\documentclass[10pt,conference]{IEEEtran}
\IEEEoverridecommandlockouts
\usepackage{amsmath,amssymb,amsfonts}
\usepackage{xcolor}
\definecolor{fixmeColor}{rgb}{0.8,0.1,0.1}

\definecolor{stringColor}{rgb}{0.8,0.1,0.1}
\usepackage{listings}
\usepackage{courier}
\usepackage[colorlinks=true, linkcolor=blue!70!black, citecolor=green!70!black, urlcolor=blue!70!black]{hyperref}
\usepackage{doi}
\definecolor{airforceblue}{rgb}{0.36, 0.54, 0.66}
\definecolor{bluegray}{rgb}{0.4, 0.6, 0.8}
\definecolor{darkcerulean}{rgb}{0.03, 0.27, 0.49}
\lstdefinelanguage{LF}{
  keywords={@absent_after, action, after, deadline, federated, implements, import, initial, input, logical, main, @maxwait, mode, msec, msecs, new, output, physical, preamble, reaction, reactor, sec, secs, shutdown, startup, state, tardy, target, time, timer, usec, usecs},
  emph={L,name, type, init, effect, instance}, emphstyle=\itshape,
  keywordstyle=\color{black}\bfseries,
  ndkeywordstyle=\color{darkcerulean}\bfseries,
  identifierstyle=\color{black},
  sensitive=false,
  comment=[l]{//},
  morecomment=[s]{/*}{*/},
  commentstyle=\color{airforceblue}\ttfamily,
  stringstyle=\color{black}\ttfamily,
  morestring=[b]',
  morestring=[b]"
}
\usepackage{float}

\usepackage{algorithmic}
\usepackage{graphicx}
\usepackage{textcomp}
\usepackage{xcolor}
\def\BibTeX{{\rm B\kern-.05em{\sc i\kern-.025em b}\kern-.08em
    T\kern-.1667em\lower.7ex\hbox{E}\kern-.125emX}}
\usepackage{cite}

\usepackage{ifthen}
\newboolean{showcomments}
\setboolean{showcomments}{true} 
\newcommand{\mynote}[3]{%
  \ifthenelse{\boolean{showcomments}}{%
   \fbox{\bfseries\sffamily\scriptsize#1}%
   {\small$\blacktriangleright$\textsf{\emph{\color{#3}{#2}}}$\blacktriangleleft$}}%
  {%
   \@bsphack
   \@esphack
}%
}

\usepackage[inkscapelatex=false]{svg}

\begin{document}

\title{Maxwait: A Generalized Mechanism for Distributed Time-Sensitive Systems
\thanks{This work was supported in part by the National Science Foundation, Award 2449200, POSE: Phase II: An Open-Source Ecosystem for Lingua Franca.}
}


\author{\IEEEauthorblockN{Francesco Paladino}
\IEEEauthorblockA{\textit{EECS} \\
\textit{UC Berkeley}\\
Berkeley, CA, USA \\
ORCID: 0009-0009-0903-1149}
\and
\IEEEauthorblockN{Shulu Li}
\IEEEauthorblockA{\textit{EECS} \\
\textit{UC Berkeley}\\
Berkeley, CA, USA \\
ORCID: 0009-0001-7289-6577}
\and
\IEEEauthorblockN{Edward A. Lee}
\IEEEauthorblockA{\textit{EECS} \\
\textit{UC Berkeley}\\
Berkeley, CA, USA \\
ORCID: 0000-0002-5663-0584}
}

\maketitle

\begin{abstract}
Distributed time-sensitive systems must balance timing requirements (availability) and consistency in the presence of communication delays and synchronization uncertainty. This paper presents maxwait, a simple coordination mechanism with surprising generality that makes these tradeoffs explicit and configurable. We demonstrate that this mechanism subsumes classical distributed system methods such as PTIDES, Chandy-and-Misra with or without null messages, Jefferson's Time-Warp, and Lamport's time-based fault detection, while enabling real-time behavior in distributed cyber-physical applications. The mechanism can also realize many commonly used distributed system patterns, including logical execution time (LET), publish and subscribe, actors, conflict-free replicated data types (CRDTs), and remote procedure calls with futures. More importantly, it adds to these mechanisms better control over timing, bounded time fault detection, and the option of making them more deterministic, all within a single semantic framework. Implemented as an extension of the Lingua Franca coordination language, maxwait enforces logical-time consistency when communication latencies are bounded and provides structured fault handling when bounds are violated.
\end{abstract}


\section{Introduction}

Distributed time-sensitive systems present unique challenges.
What we mean by ``time-sensitive'' is systems where components are considered to have \emph{failed} when timing requirements are not met.
This contrasts with timing as a performance metric, where a slow system has not failed but merely underperformed.
What we mean by ``distributed'' is software systems that run on multiple computers connected by networks that require some measure of consistency.
Independent systems running on multiple computers but not requiring consistency are not of interest in this paper.
What we mean by ``consistency'' is (intuitively) agreement on shared information.
This notion can be made precise and formal.
Here, we will adopt the definition given by Lee et al.~\cite{LeeEtAl:23:CAL_CPS}.

In this paper, we review fundamental limits that imply that maintaining consistency comes with a fundamental timing cost (unavailability) that is a function of latency, where latency is a measure of the time it takes for distributed components to communicate.
Further, we review fundamental limits showing that, in order to bound the timing cost, coordination is required.
Coordination means that the components cannot take action without some form of communication with other components.

Given these fundamental limits, every time-sensitive distributed system is a compromise, where design decisions consider the tradeoffs.
One design may put top priority on consistency, giving up timing guarantees, whereas another may put top priority on availability, allowing some measure of inconsistency.
These choices depend on the application, and a single application may make different choices for different parts.

This paper describes a mechanism that we call ``maxwait'' that proves surprisingly general in its ability to provide many of the decision points in these fundamental tradeoffs.
We give an implementation as an extension of the Lingua Franca (LF) coordination language~\cite{LohstrohEtAl:21:Towards}.
The mechanism first appeared, to our knowledge, in the form of a protocol called PTIDES~\cite{Zhao:07:PTIDES} (although it has roots in Lamport~\cite{Lamport:84:TimeStamps}).
It was later independently reinvented to become part of the backbone of Google Spanner, one of the world's largest distributed databases~\cite{CorbettEtAl:12:Spanner}.
Our contribution in this paper is to clarify and refine the semantics of the mechanism and, more importantly, to show the generality of the mechanism and its ability to control timing.
It subsumes many widely used distributed computing techniques as special cases and makes mixing these techniques in the same application much more attractive.
More importantly, our \textit{maxwait} mechanism brings timing concerns to the foreground, enabling enforcement of timing requirements and the handling of faults when enforcement becomes impossible.

The paper is organized as follows. Section~\ref{sec:limits} recalls the theoretical background of the tradeoff between consistency and availability. Section~\ref{sec:decentr-coord} introduces the \textit{maxwait} mechanism implemented in LF. Section~\ref{sec:techniques} shows with LF programs that \textit{maxwait} realizes many classical coordination strategies and enables control over timing for time-sensitive applications. Finally, Section~\ref{sec:special} further demonstrates the generality of \textit{maxwait} by showing that it can realize many commonly used communication strategies for distributed systems.

\section{Fundamental Limits}
\label{sec:limits}

Lee et al.~give the CAL theorem~\cite{LeeEtAl:23:CAL_CPS} (for consistency, availability, and latency), which is a quantified version of Brewer's well-known CAP theorem~\cite{Brewer:17:CAP} (for consistency, availability, and partition tolerance).
The CAL theorem quantifies each of the three parameters and gives a formula that relates the three.
The formula is a system of equations in a max-plus algebra~\cite{Baccelli:92:MaxPlus} that shows how increasing latency implies increasing either unavailability or inconsistency.
Latency increases when execution times or network delays increase.
Network partitioning is simply a special case in which the network delay becomes infinite.
Lee et al.~show that network delays are indistinguishable from clock synchronization errors in any one-way communication system, which means that clock synchronization errors should be considered equivalent to network degradations.

Consistency has many mutually inconsistent treatments and interpretations in the literature~\cite{Helland:21:ConGame}.
Here, we define consistency as agreement on the state of a replicated deterministic state machine. Each node maintains a copy of the state machine, and inputs to the node drive changes in the state.
In this paper, we will be interested in two forms of agreement:
\begin{enumerate}
    \item Eventual consistency: All nodes that have seen the same inputs agree on the final state of the state machine.
    \item Logical-time consistency: All state machines progress through the same sequence of states.
\end{enumerate}
These notions require elaboration.
A deterministic state machine is a tuple $(\mathbb{S}, s_0, \mathbb{I}, r)$, where $\mathbb{S}$ is a set of states, $s_0 \in \mathbb{S}$ is an initial state, $\mathbb{I}$ is an input alphabet (a set containing all allowed inputs for a single state transition), and $r\colon \mathbb{S} \times \mathbb{I} \to \mathbb{S}$ is a state transition function.
Let $\mathbb{I} ^ *$ be the set of all finite sequences $I_n = (i_1, i_2, \cdots , i_n) \in \mathbb{I}^*$ of inputs, where $i_k \in \mathbb{I}, 1 \le k \le n$.
Then we can define a function $R\colon \mathbb{S} \times \mathbb{I}^* \to \mathbb{S}$ such that $R(s_0, I_n) = r(r(\cdots r(r(s_0, i_1), i_2))\cdots,i_{n-1}), i_n ) = s_n$, where $I_n \in \mathbb{I}^*$ is an input sequence of length $n$ and $s_n$ is the state of the state machine after reacting to these inputs in order.

Given these definitions, it is easy to see that if all nodes get inputs in the same order, then logical-time consistency is assured, and eventual consistency is implied.

We call this notion ``logical-time consistency'' because each state machine replica proceeds through a sequence of states $s_0, s_1, \cdots , s_k, \cdots , s_n$, and the index $k$ can be viewed as a tick of a logical clock shared by all nodes.
I.e., each replicated state machine is in state $s_k$ at the $k^{th}$ tick of a logical clock.
We can, without loss of generality, make a closer association of physical time by defining a sequence $t_0, t_1, \cdots$ where each $t_k$ is a measure of the physical time of the $k^{th}$ tick according to some physical clock.
We can then say that each state machine is in state $s_k$ at logical time $t_k$.
We call this a ``logical time'' not a physical time because whatever clock is being used to provide the measure of physical time can only be accessible to at most one node.
For the other nodes, therefore, $t_k$ is not a measure of physical time with respect to any clock they have access to.
In particular, it is \emph{not} the time at which they make the $k^{th}$ state transition according to any local clock.

We consider any coordination mechanism that controls the ordering of distributed events to be essentially providing a model of time.
It provides Lamport's ``happens before'' relation~\cite{Lamport:78:Time}.
In a distributed system, it is a partial order relation, in which unrelated events have no relative ordering.
Such a model of time is consistent with modern physical notions of time.
Our logical-time consistency framework is similar to Lamport clocks and to the model of time in synchronous languages~\cite{Benveniste:91:Synchronous}.

In all variants of consistency, inconsistencies can arise when events are handled in different orders across a distributed system, and when the handling of events is sensitive to the order in which they are handled.
For example, if two nodes update a local copy of a database record in different orders, they will disagree on the final value of the record.
One way to ensure eventual consistency is to ensure that handling events in different orders has no effect on the final outcome.
We call this family of techniques ``coordination-free'' because they eliminate the need for coordination.
These techniques will not achieve logical-time consistency but can achieve eventual consistency.
Another way to ensure consistency is to control ordering, which requires some form of coordination.
Such techniques make logical-time consistency possible.

We can define coordination more formally and relate the two forms of consistency under differing coordination strategies.
Consider a collection of messages that are to be provided to a collection of nodes, each maintaining a replicated state machine.
These messages must be presented in some way and some order to each state machine, but the messages come from diverse parts of the network, so they are not initially ordered.
Let $\mathbb{N}$ be the set of natural numbers, and $M \in \mathbb{N}^\mathbb{I}$ be a multiset representing the totality of messages to be presented to each state machine. $M$ is a function that maps each message in $\mathbb{I}$ to the number of times the message is presented to the state machine.
Define a \textbf{coordination function} $\mathcal{C}\colon \mathbb{N}^\mathbb{I} \to 2^{(\mathbb{I}^*)}$.
Given a collection of messages $M \in \mathbb{N}^\mathbb{I}$, $\mathcal{C}(M)$ yields a set of allowed sequences that may be presented to the state machines.

The \textbf{trivial coordination function} is one that allows the messages in $M$ to be presented in any order to each state machine. Formally, $\mathcal{C}(M) = \text{Perm}(M)$, where Perm is the permutation function:
\begin{align*}
\text{Perm}(M) &= \{ (i_1, \cdots , i_{|M|}) \in \mathbb{I}^{|M|} \\
&|~\forall i \in \mathbb{I}, |\{k : i_k = i \}| = M(i) \}.
\end{align*}
The trivial coordination function does not perform coordination at all, as the only constraint on the messages is that they need to appear in each sequence the number of times prescribed by $M$.
To achieve eventual consistency with such a coordination function will require severe constraints on the state machine structure (see the discussion below of ACID, CRDTs, and CALM for ways of achieving this).

The \textbf{logical time coordination function} assumes that $\mathbb{I}$ has the form $\mathbb{I} = \mathbb{T} \times \mathbb{D}$, where $\mathbb{T}$ is a totally ordered set of \textbf{timestamps} and $\mathbb{D}$ is the set of all possible data payloads.
It then maps any multiset $M \in \mathbb{N}^\mathbb{I}$ into a sequence of $(t,d)$ pairs such that the timestamps are strictly increasing.
Such a coordination function that presents each state machine with the messages in the same set $M$ will present them in the same order, and hence will achieve logical time consistency and eventual consistency.

In this paper, we describe the \textbf{maxwait coordination mechanism}, which realizes the logical time coordination function precisely if latencies are bounded by specified bounds.
And it provides exception handlers for detecting violations of this coordination function when these bounds are exceeded.
It benefits from clock synchronization and reliable, in-order message delivery (e.g. TCP/IP).
With just these two coordination mechanisms, and no more, we can realize a rich variety of distributed computing patterns that are used in practice.
These include techniques that achieve logical-time consistency, those that only achieve eventual consistency, and, more interestingly, those that relax consistency to improve availability.
This latter category is what enables real-time distributed computing.

\section{Decentralized Coordination}
\label{sec:decentr-coord}

One way to eliminate potential inconsistencies is to make the nodes of a distributed system process events in the same order, thus guaranteeing logical-time consistency. In this section, we will focus on an approach to control event ordering that does not rely on a centralized entity. Rather, each node makes its own decision on when it is reasonable to make progress by using only the inputs from other nodes and its physical clock. Hence, the name ``decentralized coordination.'' Clock synchronization techniques then become one of the building blocks of this strategy; fortunately, they have been improving steadily and are starting to become ubiquitous~\cite{GengEtAl:18:ClockSync,EidsonStanton:15:Time,Mills:06:NTP,Eidson:06:1588}.

We have implemented the decentralized coordination mechanism in Lingua Franca (LF)~\cite{LohstrohEtAl:21:Towards}, an open-source coordination language based on reactive components called \textit{reactors}. Reactors react to events by invoking \textit{reactions} that implement the component behavior. Reactors communicate through ports by exchanging messages. Messages carry a generalized timestamp called a \textit{tag} that enables event ordering, helping to achieve consistency even in distributed settings. Timestamps realize the abstraction of logical time, distinct from the physical time measured on the specific hardware platform on which the application is running, as explained in the previous section.

LF supports the design of distributed applications through the concept of \textit{federation}, where each distributed node is called \textit{federate}. The runtime provides an abstraction of the remote communication functionality to allow federates to exchange messages. LF features a clock synchronization algorithm~\cite{GengEtAl:18:ClockSync} to align the clock of the federates. Alternatively, this can be disabled and replaced by external synchronization algorithms, such as NTP (the network time protocol)~\cite{Mills:06:NTP} or PTP (the precision time protocol)~\cite{Eidson:06:1588}.

The decentralized coordination technique implements a coordination function $\mathcal{C}$ that ensures logical-time consistency under clearly stated assumptions about network behavior. Each federate will process messages in strictly increasing tag order, even when these come from different federates.

Each federate has zero or more input ports and output ports. Output ports produce a stream of messages with strictly increasing tags and send them to one or more input ports of other federates. Each input port receives messages from at most one output port.
These messages are assumed to be received reliably in tag order, as realized, for example, by using a TCP socket for communication.
Of course, no communication protocol achieves perfect in-order reliable delivery, so when this requirement fails, there will be
detectable faults.


The coordination takes effect when a federate receives messages on its input ports. Each federate has a \textit{current tag} attribute to keep track of the messages it has already processed. Each port has a \textit{last known tag} property that stores the tag of the last message received on that port. The \textit{last known tag} is monotonically increasing (absent faults). The in-order property holds for a single federate-to-federate connection; when a federate receives input from multiple sources, coordination is necessary to guarantee unambiguous event processing.

Consider a federated LF program $F$. Each federate $f \in F$ has $n_f$ input ports in the set $P_f = \{p^f_1, p^f_2, ..., p^f_{n_f}\}$. Each input port $p^f_k$ has its last known tag property $\text{\textit{last}}_{p^f_k}$.
Suppose the federate $f$ has an event with tag $t$ that, as far as it knows, is the next event to process.
This could be a locally generated event or a message received with tag $t$ on some port.
To process this event, the federate $f$ must advance its current tag $t^{\text{\textit{curr}}}_f$ to $t$. The advancement of the tag is governed by decentralized coordination and is controlled by the \textit{maxwait} parameter, which is a property of the federate $f$, denoted \textit{maxwait}$_f$. Formally, $f$ can advance its current tag $t^{curr}_f$ to $t$ when one of the two following conditions holds:
\begin{subequations}
\begin{align}
  \forall p^f_l \in P_f,\; \quad \text{\textit{last}}_{p^f_l} \geq t \label{eq:all_inputs_known} \\
  T_f \geq t \;+\; \text{\textit{maxwait}}_f \label{eq:maxwait_firing},
\end{align}
\end{subequations}
where $T_f$ is the reading of $f$'s local clock measuring physical time.
Eq.~(\ref{eq:all_inputs_known}) states that the federate $f$ can advance to tag $t$ if all input ports have last known tag greater than or equal to tag $t$. We say concisely that ``all input ports are \textit{known} up to and \textit{including} tag $t$.'' Because of the in-order delivery of messages, before moving to tag $t$, all ports have received all messages with earlier and equal tags, and thus the federate $f$ has processed all events with earlier tags. If (\ref{eq:all_inputs_known}) is always satisfied, logical-time consistency is assured.

Eq.~(\ref{eq:maxwait_firing}) provides a timeout.
This condition kicks in when the local physical clock reports a time $T_f$ that is greater than the timestamp $t$ by at least \textit{maxwait}$_f$.
If \textit{maxwait}$_f$ is finite, it provides an upper bound on the time that the federate will wait for ports to become known.
If when this bound is reached there are still ports that are not known up to tag $t_k$, $f$ assumes that those ports will not receive any messages with tags less than or equal to $t_k$.
Based on this assumption, $f$ proceeds to update its current tag to $t_k$ and process any inputs it has received with that tag.

When condition (\ref{eq:maxwait_firing}) occurs, logical time consistency is assured only if the assumption made by $f$ is correct.
If we have an upper bound on the \textit{apparent latency}, setting \textit{maxwait} to this bound ensures logical-time consistency.
The apparent latency is defined by Lee et al.~\cite{LeeEtAl:23:CAL_CPS} to be the difference between the timestamp of a message and the physical time it is received.
It is the sum of network latency, clock synchronization error, and any computation overhead incurred by the construction and conveyance of the message.

In most practical systems, it is difficult to guarantee a bound on apparent latency, but for many real-time systems, conservative estimates may be reasonable.
If the application is built on a time-sensitive network (TSN), for example, then we may be able to assume tight bounds on apparent latency.
Nevertheless, violations may occur, in which case a federate may receive a message with tag $t$ after it has advanced to $t$ or beyond.
Such a message is said to be \textit{tardy}.
This is a critical condition because it means that events will not be handled in tag order.
Such a condition is called a \textit{safe-to-process} (STP) violation. Tardy messages are processed specially; the federate invokes a fault handler (an STP violation handler) at the earliest possible tag.


\section{Coordination Techniques for Consistency}
\label{sec:techniques}

The consistency problem in distributed systems has been widely investigated in the past.
Our contribution here is to show that our \textit{maxwait} coordination mechanism realizes many of these classical techniques as special cases and, more importantly, enables explicit control over timing.
We demonstrate this claim with Lingua Franca programs that realize conservative techniques, which enforce consistency at the expense of timing, coordination-free techniques, which forgo control over timing except in special circumstances, and optimistic techniques, which can suffer large timing penalties.
We then show that we can gain explicit control over timing by balancing the requirement for consistency against real-time requirements.
Finally, we show that the \textit{maxwait} mechanism can detect and react to faults in bounded time.
All strategies are described using simple, realistic examples.

\subsection{Conservative Techniques without Null Messages}
\label{sec:conservative}


Conservative techniques have been widely used to guarantee in-order event processing in distributed systems. One of the first examples dates back to Chandy and Misra~\cite{ChandyMisra:88:DDE}, who designed a distributed discrete-event simulation technique. Although it was intended for simulation, not for system implementation, it is relevant because it provides a mechanism for ensuring that all nodes process events in timestamp order.
As with our \textit{maxwait} mechanism, Chandy and Misra assume that each communication channel between nodes provides reliable in-order delivery of messages.
It then applies the policy given in Eq.~\ref{eq:all_inputs_known} above, where it waits for inputs from both remote nodes before processing the one with the lesser timestamp.
Eq.~\ref{eq:maxwait_firing} becomes irrelevant.

We achieve Chandy and Misra's technique simply by setting \textit{maxwait} to \textbf{forever}.
To limit the waiting time, Chandy and Misra propose that the remote nodes send periodic ``null messages,'' timestamped messages that indicate to the receiving node that no messages with the specified timestamp or less are forthcoming.
We will show how such null messages can be easily added in LF and can be fine tuned to achieve specific timing requirements.

If network latencies are not bounded, then even with periodic null messages, a receiving node may become blocked for an indefinite period of time.
For some applications, this is the appropriate real-time behavior. We illustrate this with an automated aircraft door example from Lin and Lee~\cite{LinLee:25:Models}, in which unbounded wait is the wisest coordination approach.

Aircraft doors on passenger flights are currently managed manually by flight attendants. Before takeoff, the flight attendants \textit{arm} the door; if the door is opened in this state, an evacuation slide is automatically inflated and deployed for emergency evacuation. When the aircraft is at a gate, before opening the door, the flight attendants disarm it to avoid deployment of the evacuation slide. Flight attendants disarm the door only when they see through the porthole the ramp that will allow the passengers to disembark the aircraft.

Suppose we wish to provide a service to manage the door.
This service will be implemented on a networked node that provides a \textit{disarm} service and an \textit{open} service.
Suppose that another service, implemented on another node, provides a computer-vision system that checks for a ramp at the door and issues either a \textit{true} or \textit{false} to the \textit{disarm} service, depending on whether a ramp is present.
When the door service receives an \textit{open} request, what should it do?

\begin{figure}[t]
  \centering
	  \includegraphics[scale=0.595]{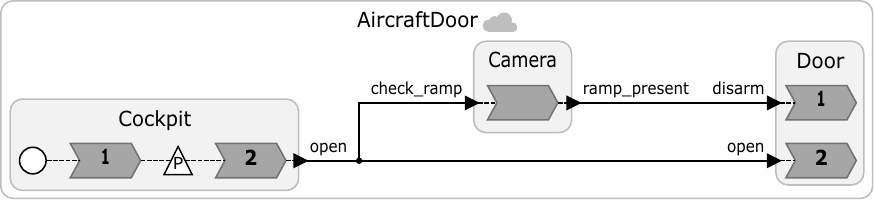}
  \caption{Diagram of a Lingua Franca application to open an aircraft door remotely.}
  \label{fig:aircraft-door}
\end{figure}

Fig.~\ref{fig:aircraft-door} shows the implementation in LF of a system to remotely open an aircraft door in the armed state.
Such diagrams are automatically generated by the LF tools from the source code for the program.
Each rounded box in the diagram is a reactor, and the chevrons are the reactions that react to events and implement the reactor behavior. The application is federated, so every reactor within the top-level AircraftDoor reactor is a federate that potentially runs on a different physical host. Communication between federates is implemented via TCP/IP, which provides reliable, in-order delivery.

The door implements two independent remote services, door disarming and door opening, encoded by two different reactions in the Door reactor. Suppose that the pilot in the cockpit issues a command to open the door.
As shown in Fig.~\ref{fig:aircraft-door}, this command triggers a camera that will verify the presence of a ramp. The camera notifies the disarming service of whether a ramp is present.


The purpose of the system is to open the door in reaction to the command from the cockpit whether or not a ramp is present. If a ramp is present, it is imperative that the door be disarmed before being opened. Hence, the door, upon receiving the open command from the cockpit, should wait for input from the camera before opening.

The order in which messages are processed is crucial in this application. When the disarm and open commands arrive with the same tag, the disarm service needs to be invoked before opening the door, otherwise the escape slide will be erroneously deployed. LF guarantees determinism in the execution order of reactions with logically simultaneous inputs, and the order is given by the order of declaration of the reactions inside the reactor. It is then sufficient to declare the disarm reaction before the open one. The diagram confirms the execution order by labeling the disarm reaction with 1 and the open reaction with 2.

The problem is that even though the messages are logically simultaneous, they do not arrive at the same physical time. The open command from the cockpit is likely to arrive before the clearance from the camera because the camera realizes an expensive computer vision algorithm. The door, consequently, has to wait for both inputs before invoking the opening service.

Using LF and the decentralized coordination, the wait for inputs can be customized using the \textit{maxwait} attribute detailed in Sec.~\ref{sec:decentr-coord}. This specific application requires \textit{maxwait} set to \textit{forever}, an LF keyword that means infinite time. In fact, the Door reactor cannot safely proceed without receiving a verdict from the Camera reactor, so it waits indefinitely for that input before processing the door open input.

Setting \textit{maxwait} to \textit{forever} implies an unbounded wait for inputs, which is how conservative techniques like Chandy and Misra~\cite{ChandyMisra:88:DDE} achieve consistency. Unbounded wait times create concerns if the communication latency becomes large for some reason, such as a network or host failure. In these cases, the application might block indefinitely waiting for inputs. However, the aircraft door application is an example in which it is better to be stuck than to act based on inconsistent information: opening the door without a response from the camera would cause severe problems. Looking at this scenario from the perspective of the CAL theorem~\cite{LeeEtAl:23:CAL_CPS}, consistency trumps availability: when the latency increases, availability should be sacrificed, not consistency.
A reasonable real-time constraint for this application is not a deadline on opening the door (unsafely), but rather a deadline on detecting network or node failures that prevent the disarm message from arriving.
We will show in Sec.~\ref{sec:fault} how to detect such failures in bounded time.

\begin{figure}
\begin{lstlisting}[language=LF,escapechar=|]
target C { |\label{lin:target-lang}|
  coordination: decentralized,
  keepalive: true
}
import Cockpit from "lib/Cockpit.lf"
import Camera from "lib/Camera.lf"

reactor Door { |\label{lin:door}|
  input open: bool
  input disarm: bool
  state isDisarmed: bool = false |\label{lin:disarmed}|
  state isOpen: bool = false |\label{lin:opened}|

  reaction(disarm) {= |\label{lin:disarm}|
    if (disarm->value && !self->isDisarmed) {
      self->isDisarmed = true;
      printf("Door disarmed\n");
    } else if (!disarm->value && self->isDisarmed) {
      self->isDisarmed = false;
      printf("Door armed\n");
    }
  =}
  reaction(open) {= |\label{lin:open}|
    if (open->value && !self->isOpen) {
      self->isOpen = true;
      printf("Door open\n");
    } else if (!open->value && self->isOpen) {
      self->isOpen = false;
      printf("Door closed\n");
    }
  =}
}
federated reactor { |\label{lin:federated}|
  c = new Cockpit()
  v = new Camera()
  @maxwait(forever) |\label{lin:maxwait}|
  d = new Door()
  c.open -> d.open |\label{lin:connections-start}|
  c.open -> v.check_ramp
  v.ramp_present -> d.disarm |\label{lin:connections-end}|
}
\end{lstlisting}
\caption{Lingua Franca code of the aircraft door example.}
\label{fig:aircraft-door-code}
\end{figure}

Fig.~\ref{fig:aircraft-door-code} shows the code implementing the aircraft door application, with the focus on the Door reactor. LF is a \emph{coordination} language, where the reactor behavior is specified using one of the supported conventional programming languages: C, C++, Python, TypeScript, and Rust. This example is written in C, as indicated by the directive on line~\ref{lin:target-lang}.

Line~\ref{lin:federated} defines the main reactor that instantiates all reactors within its body: Cockpit, Camera, and Door. The \texttt{federated} keyword indicates that the application is federated, that is, distributed. Lines~\ref{lin:connections-start}-\ref{lin:connections-end} establish connections between reactors through ports, as shown in the diagram of Fig.~\ref{fig:aircraft-door}.

Line~\ref{lin:door} defines the Door reactor. The definition begins with input and output ports and state variables, followed by reactions. Both ports and variables have types, boolean in this case. The reaction on line~\ref{lin:disarm} reacts to any input received on the \texttt{disarm} port and will disarm or arm the door. The reaction body is delimited by the fences \texttt{\{=} ... \texttt{=\}}, which wrap code written in the target language (C in our case). The \texttt{isDisarmed} state variable on line~\ref{lin:disarmed} maintains the current state of the door and is updated in the reaction body according to the command from the camera. Similarly, the reaction on line~\ref{lin:open} reacts to the open command sent by the Cockpit reactor and updates the \texttt{isOpen} state variable accordingly.

As described in Sec.~\ref{sec:decentr-coord}, \textit{maxwait} for decentralized coordination is a federate property. This is set to \texttt{forever} in the main reactor on line~\ref{lin:maxwait} with an annotation right before the federate instantiation. 
Because of this setting, the Door reactor will not react to an input on either port until it has received an input on the other port with a greater or equal timestamp.
This is exactly what we need for safe behavior in this case.


The ``null messages'' of Chandy and Misra are not needed in this application because the door cannot safely advance when either of its two inputs is missing.
We consider next an example that requires null messages.

\subsection{Conservative Techniques with Null Messages}
\label{sec:conservative-null}

Consider an example from Lee~\cite{Lee:25:Actors} of a simple banking application that maintains a distributed, replicated account balance. Customers can deposit and withdraw money through ATMs.
We assume a real-time requirement that the ATMs respond within specified deadlines.


\begin{figure}[b!]
\begin{center}
    	  \includegraphics[scale=0.5]{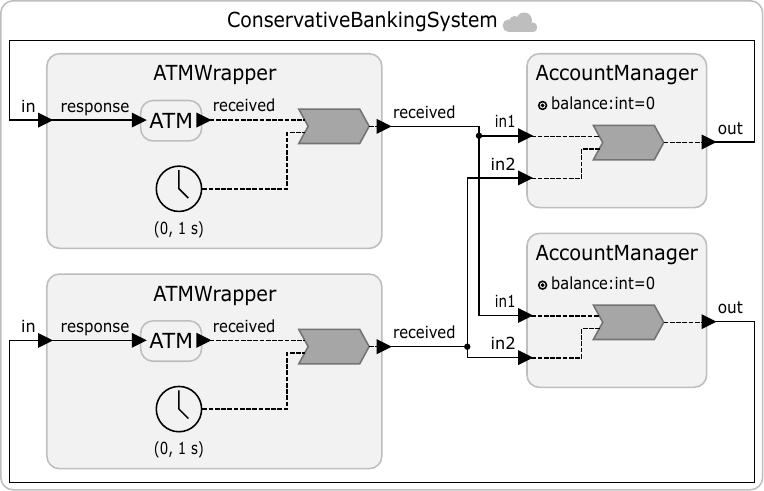}
\end{center}
\begin{lstlisting}[language=LF,escapechar=|]
import ATM from "lib/ATM.lf"
reactor ATMWrapper(..., null_msg_t: time = 1s) {
  input in: int
  output received: int
  timer t(0, null_msg_t) |\label{lin:timer}|
  w = new ATM( ... )
  in -> w.response
  reaction (w.received, t) -> received {= |\label{lin:null-msg}|
    if (w.received->is_present) {
      lf_set(received, w.received->value);
    } else {
      lf_set(received, 0); // Send null message
    }
  =}
}
reactor AccountManager() {
  input in1: int
  input in2: int
  output out: int
  state balance: int = 0 |\label{ln:state}|
  reaction(in1, in2) -> out {= |\label{ln:reactiont}|
    int in1_val = 0;
    int in2_val = 0;
    if (in1->is_present) {
      if (self->balance >= -in1->value) { |\label{lin:overdraft-check-1}|
        self->balance += in1->value;
      } else {
        self->balance -= 30; // Apply penalty
      }
    }
    if (in2->is_present) {
      if (self->balance >= -in2->value) { |\label{lin:overdraft-check-2}|
        self->balance += in2->value;
      } else {
        self->balance -= 30; // Apply penalty
      }
    }
    lf_set(out, self->balance);
  =}
}
federated reactor {
  w1 = new ATMWrapper( ... )
  w2 = new ATMWrapper( ... )
  @maxwait(forever) |\label{lin:forever-penalty-1}|
  a1 = new AccountManager()
  @maxwait(forever) |\label{lin:forever-penalty-2}|
  a2 = new AccountManager()
  w1.received -> a1.in1
  w2.received -> a2.in2
  w1.received -> a2.in1
  w2.received -> a1.in2
  a1.out -> w1.in
  a2.out -> w2.in
}
\end{lstlisting}
\caption{A conservative account manager with overdraft penalty.} \label{fig:conservative-bank-code}
\end{figure}

An LF realization of a small version of this banking application is shown in Fig.~\ref{fig:conservative-bank-code}.
This banking system is distributed; two account managers, potentially in physically different places, need to eventually agree on the account balance.
Each maintains a copy of the balance in a state variable declared on line \ref{ln:state}.

Two distinct ATMs allow bank transactions.
The ATM implementation details are omitted for brevity, but our prototype models the ATM using a web page that connects to the LF program via a web socket. Customers enter the amount to be deposited or withdrawn on a web page. Customer requests are positive integers for deposits and negative integers for withdrawals.
Line \ref{lin:timer} defines a timer with an offset of zero and period of one second, shown in the diagram with the clock icon and the (offset, period) parameter values below the icon.
The offset is relative to the start time of the program.
The reaction starting on line \ref{lin:null-msg} periodically sends null messages, which are just zero-valued requests.

Requests are forwarded to the AccountManager reactors, which update the account balance accordingly. The AccountManager reactor sends the updated balance back to the ATM reactor as a response.
Lines \ref{lin:forever-penalty-1} and \ref{lin:forever-penalty-2} ensure that both account managers process inputs in timestamp order, thus giving timestamp consistency and eventual consistency.

In this realization, the reaction starting on line \ref{ln:reactiont} executes when either input receives a timestamped message.
Based on the local copy of the balance, transactions will be either allowed or disallowed, and, in the latter case, a penalty of \$30 will be imposed.
This is a particular business logic, one of many that you could implement.
The two account managers need to agree on which requests are allowed and disallowed.
Hence, when they process a timestamped request, they need to agree on the balance at that timestamp.

In this realization, the two account managers will agree on the balance at all timestamps.
However, there could be severe real-time penalties.
A delay in responding to requests will be incurred that depends on the period of the null messages and the network latency, making a deadline difficult to assure.
Moreover, network failures will render the system unresponsive.
We next explore techniques that relax coordination and remove these dependencies.

\subsection{ACID 2.0, CRDTs, and CALM}
\label{sec:acid}

It is possible to reduce or eliminate the sensitivity to ordering using any of a family of techniques such as ACID merge, CRDTs, and logically monotonic merge.
These are ``coordination-free'' techniques.
ACID, in this usage, stands for Associative, Commutative, Idempotent, and Distributed, and was introduced by Helland and Campbell~\cite{HellandCampbell:09:ACID2}.
In distributed systems, idempotence means that if a message is delivered more than once, it has the same effect as if it had been delivered exactly once. The TCP connections that we use guarantee exactly once delivery.
If distributed events are handled using ACID operations, then the order in which they are handled will not change the final result.

Shapiro et al.~introduce conflict-free replicated data types~\cite{Shapiro:11:CRDT}.
A CRDT is any data type for which updates to an instance of the data type in different orders yield the same final value.
Hellerstein et al. showed that any update to a data structure that is monotonic in a suitably chosen partial order has the same property as a CRDT.
They call this the CALM theorem, for consistency as logical monotonicity.

Li and Lee define strong eventual consistency to mean that any two nodes that see the same events yield the same final values, regardless of the order in which they process the events~\cite{LiLee:25:Consistency}.
They prove that the ACID properties are not only sufficient to achieve this, but also necessary, thereby showing that ACID and CRDTs are  equivalent while CALM is strongly related.

\begin{figure}[t]
\begin{center}
    \includegraphics[scale=0.595]{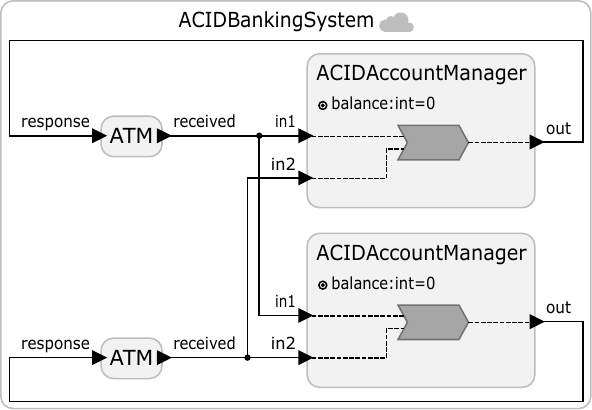}
\end{center}
\begin{lstlisting}[language=LF,escapechar=|]
import ATM from "lib/ATM.lf"

reactor ACIDAccountManager() { |\label{lin:acid-account}|
  input in1: int
  input in2: int
  output out: int
  state balance: int = 0
  reaction(in1, in2) -> out {= |\label{lin:two-input-ports}|
    if (in1->is_present) { |\label{lin:bank-logic-start}|
      self->balance += in1->value;
    }
    if (in2->is_present) {
      self->balance += in2->value;
    }
    lf_set(out, self->balance); |\label{lin:bank-logic-end}|
  =} tardy |\label{lin:tardy-empty}|
}
federated reactor { |\label{lin:federated-acid}|
  w1 = new ATM(...)
  w2 = new ATM(...)
  @maxwait(0) |\label{lin:0maxwait-1}|
  a1 = new ACIDAccountManager()
  @maxwait(0) |\label{lin:0maxwait-2}|
  a2 = new ACIDAccountManager()

  w1.received -> a1.in1
  w2.received -> a2.in2
  w1.received -> a2.in1
  w2.received -> a1.in2
  a1.out -> w1.response
  a2.out -> w2.response
}
\end{lstlisting}
\caption{Lingua Franca realization of the ACID banking system example.}
\label{fig:acid-bank}
\end{figure}

Fig.~\ref{fig:acid-bank} shows a variant of the banking example, where the account manager satisfies the ACID properties. 
Unlike the aircraft door and previous account manager examples, in this case, it \textit{is} safe to proceed without coordination because the balance update performed by the reaction in the ACIDAccountManager is associative and commutative, assuming no overflow (the TCP connection provides idempotence).
To realize the coordination-free mechanism, we set \textit{maxwait} to zero on lines~\ref{lin:0maxwait-1} and~\ref{lin:0maxwait-2}; requests sent to the ACIDAccountManager reactors will be processed as soon as they arrive. With $\text{\textit{maxwait}}=0$, STP violations may occur; to ignore them, we declare on line~\ref{lin:tardy-empty} an empty \texttt{tardy} handler.
This LF syntax tells the execution engine to invoke the ordinary reaction even on tardy inputs.
The ACID properties mean that regardless of network latencies and nondeterministic message ordering, as long as both account managers eventually see the same messages, they will eventually agree on the balance.
And meeting deadlines is easy because there is no need to wait for remote events.

However, to achieve ACID, the business logic encoded by the reaction on lines~\ref{lin:bank-logic-start}-\ref{lin:bank-logic-end} \textit{always} allows withdrawals, even when they cause an overdraft. 
This realizes a rather poor business model that will likely lead to bankruptcy.

There is a trivial way to achieve ACID properties while using the original non-ACID account manager from Fig.~\ref{fig:conservative-bank-code}.
Instead of processing events immediately, the account manager could simply collect timestamped messages in a set.
Adding elements to a set is an associate, commutative, and idempotent operation.
To determine the bank balance, however, each account manager would have to assume it has received all transactions before processing any of them.
This would obviously incur an unacceptable (infinite) timing penalty.

So far, we have shown that our \textit{maxwait} mechanism can realize both Chandy and Misra-style conservative coordination and coordination-free techniques such as CRDTs. However, neither of these classes of methods yields a good solution for our banking application. We next explore optimistic techniques.

\subsection{Optimistic}

Jefferson~\cite{Jefferson:85:TimeWarp} introduced a distributed system technique that he called ``TimeWarp,'' in which events are processed speculatively when received.
If a node later receives an event that is out of order, then the node rolls back its state to a prior ``known good'' state and reprocesses all events received received after that state.

In the context of cyber-physical systems with timing sensitivity, rollback has limitations.
Actuation in the physical world may be impossible to roll back.
In the banking example, if cash has been dispensed, it may become impossible to reverse.
Nevertheless, optimistic techniques may be useful to realize a certain business logic.

\begin{figure}[t]
  \centering
	  \includegraphics[width=\columnwidth]{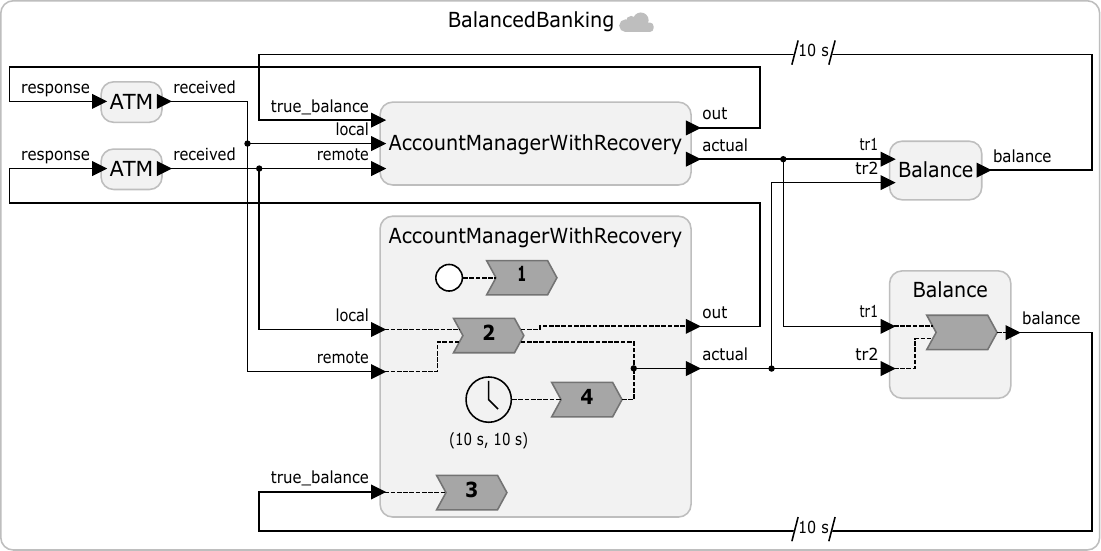}
  \caption{Lingua Franca banking system with rollback.}
  \label{fig:optimistic-bank}
\end{figure}

Fig.~\ref{fig:optimistic-bank} shows a banking system that combines conservative techniques with optimistic techniques with rollback. The ATM reactors send requested deposits or withdrawals to the AccountManagerWithRecovery reactors.
Those reactors are designed to respond within about 30 ms regardless of network conditions;
their \textit{maxwait} is set to 30 ms.
They use local estimates of the current balance to make decisions about whether to grant a withdrawal.
For example, they could grant a withdrawal even without consistency assurance if the amount of the withdrawal is small enough that the risk is modest.

Each AccountManagerWithRecovery forwards its actual transactions (deposits and withdrawals and/or penalties) to a local and a remote Balance reactor.
The Balance reactors have \textit{maxwait} set to \textit{forever}, so they maintain a timestamp consistent true balance.
Each time they update that balance, they feed it back to the local AccountManagerWithRecovery, which updates the local balance accordingly. The connections with the account managers have a \textit{logical delay} of 10 s; if a Balance reactor sends a balance update to an account manager with tag $t$, the account manager will receive that update with tag $t+10\;\text{s}$. The account managers receive the timestamp consistent balance that is ten seconds old, so this parameter is the upper bound to the duration of inconsistencies in the application.

The ten second delay is specified in LF using the syntax:
\begin{lstlisting}[numbers=none]
  b.balance -> a1.true_balance after 10 s
\end{lstlisting}
The reason for this logical delay is to give an additional ten seconds for the true balance to be formed and propagated back to the account managers.
Without it, the \textit{maxwait} of 30 ms at the account managers would likely lead to an STP violation.
As long as the total latencies stay below 10 s, no such STP violations will occur, and the account managers will never be working with true balances older than 20 s.
It is 20 s, not 10 s because the AccountManagerWithRecovery reactors also send null messages with a ten second period to ensure that the timestamp consistent statement is updated at least once every ten seconds.

The absence of a timely true balance message can be detected in reaction 4 in Fig.~\ref{fig:optimistic-bank}.
This reaction can implement any suitable business decision, including taking the ATM offline due to network failures or continuing to operate with knowledge of an increased inconsistency.

This technique for the banking system bounds unavailability to 30 ms and transient inconsistencies to 20 seconds, assuming network latencies are bounded by 10 s, thus implementing a business logic that trades off strong consistency for availability and quick response to the user.
According to the CAL theorem, there is no implementation that can bound both unavailability and inconsistency unless latency is also bounded.

\subsection{Bounded ``maxwait''}
As shown in the previous section, a finite \textit{maxwait} value is the enabler for a tradeoff between consistency and availability. The use of this technique paves the way for the development of a broad range of distributed applications having stringent timing requirements for the completion of their tasks. It is the case of cyber-physical systems, and specifically of real-time systems, in which the actuation on the external environment is only effective if it happens within a certain amount of time called a \textit{deadline}.

\begin{figure}[t]
  \centering
	  \includegraphics[scale=0.527]{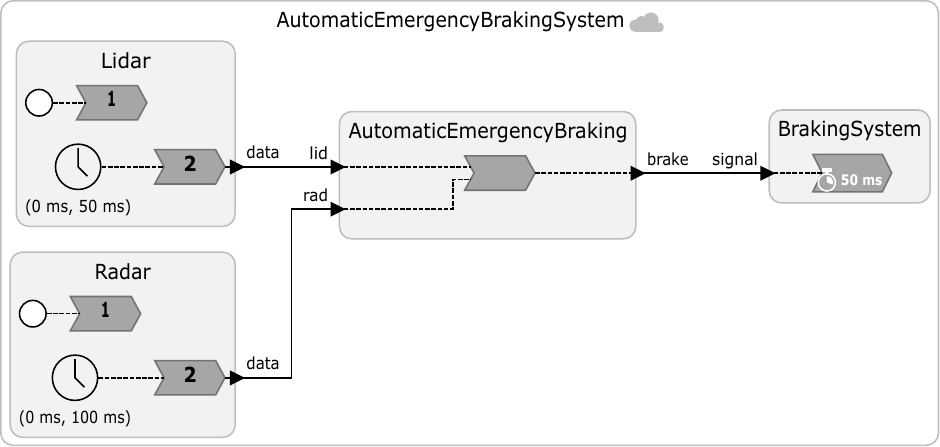}
  \caption{Lingua Franca automatic emergency braking system.}
  \label{fig:aeb}
\end{figure}

Consider an automatic emergency braking system, one of the most critical ADAS systems (advanced driver assistance system) with which modern cars are equipped. 
Fig.~\ref{fig:aeb} shows an LF program where the controller system is modeled by the AutomaticEmergencyBraking reactor, which receives data from two sensors, a lidar and a radar and uses both to detect objects or pedestrians that cross the trajectory of the car. This is a sensor fusion problem, where a diversity of sensors is used to get better reliability. When the AutomaticEmergencyBraking reactor detects a close object using the sensor data, it triggers the brake to stop the car.

Braking on time is critical to avoid crashing. The reaction of the BrakingSystem reactor has a deadline of 50 ms, which is shown in the diagram with a stopwatch icon inside the chevron. A deadline in Lingua Franca constrains the physical time at which a reaction is invoked: this time minus the logical time of its trigger should not exceed the time specified by the deadline. When it does, a deadline violation handler is invoked instead of the reaction. Deadlines also guide the Lingua Franca scheduler to prioritize the most urgent reactions. In our example, we require braking to initiate within 50 ms of detecting an obstacle or pedestrian.

The sensors are modeled with their own timer that triggers the generation of data. The lidar has a sampling frequency that is twice that of the radar: the lidar timer has a period of 50 ms, while that of the radar 100 ms. The clocks of all federates are automatically synchronized by the clock synchronization algorithm of the LF runtime (unless this is disabled). Typically, in a real use case of this kind, the clocks of sensor devices cannot be controlled by LF, but a way to work around this limitation is to resample the data collected by sensors with the timing given by a clock that the runtime can control. The sensor reactors of our application are then modeling this resampling of sensor data so that alignment of data from the two sensors is well defined.

Availability is a crucial property of this application, because we want the automatic emergency braking system to brake as fast as possible when a close object is detected. Consistency is also necessary, as sensor fusion happens with sensor data produced at the same logical time. Sensor fusion helps rule out false positives that may induce unnecessary braking. False positives may be caused by the weaknesses of the specific sensor, for example, adverse weather conditions. The key concept is to gather data produced at the same logical time by all sensors and combine them to have a more accurate estimate of possible collisions. Consistency and in-order data processing are then required.

Since the AutomaticEmergencyBraking reactor has multirate inputs, it performs two different detection strategies. At logical times 0 ms, 100 ms, 200 ms, etc, the controller expects to receive both lidar and radar data, so it invokes a sensor fusion algorithm for a more accurate obstacle prediction. Coordination is required to wait for both inputs before processing. A reasonable \textit{maxwait} value in this case is 50 ms, that is, the smallest period between the two sensors, because longer waits might cause queuing of lidar data on the input port of the controller.

At logical times 50 ms, 150 ms, 250 ms, etc, only the lidar sensor produces new data, so no coordination is needed. The \textit{maxwait} in this case can be zero because only a single input is expected and no other events can cause logical time to advance.
Before that input arrives, there is nothing to do;
once the input arrives, there is no need to wait for a second input because only that one input is expected.

In summary, consistency for sensor fusion suggests using $\textit{maxwait} = 50\;\text{ms}$ when inputs from both sensors are expected, while availability suggests $\textit{maxwait} = 0$ when only the lidar input is expected. The two values are at odds, and any value in between would mean sacrificing both properties at the same time.

To support such examples, we have extended LF to support dynamically changing the \textit{maxwait} in a reaction body using an \texttt{lf\_set\_fed\_maxwait} API function, which takes as input parameter the new \textit{maxwait} value to set.
Using this, it is easy to alternate between a \textit{maxwait} of 50 ms and one of zero.
\subsection{Fault Detection and Handling}\label{sec:fault}


In the previous section, we proposed that the AutomaticEmergencyBraking reactor of Fig.~\ref{fig:aeb} could alternate between a \textit{maxwait} of 50 ms (when two inputs are expected) and a \textit{maxwait} of zero (when only the lidar input is expected).
The choices, however, could lead to STP violations.
Suppose, for example, that the lidar data, the only expected, is delayed sufficiently that the next radar input arrives before the lidar input.
In that case, because the \textit{maxwait} is zero, the radar input will be handled immediately, and logical time will be advanced to the timestamp of the radar input.
When the lidar input eventually arrives, it will be tardy, having a timestamp earlier than that of the previously processed radar input.

Such an STP violation can be detected and handled using a syntax like this:
\begin{lstlisting}[numbers=none]
  reaction(lid, rad) -> brake {=
    // Normal reaction
  =} tardy {=
    // Fault handler
  =}
\end{lstlisting}
The tardy handler can choose what to do with late sensor data.
For example, it could simply discard the data if tardiness is infrequent, but issue a maintenance alert and disable the ADAS system if it is frequent.

However, the tardy handler is only invoked when a message arrives with larger-than-expected latency.
What if we need fault handling to occur sooner?

Because the example is based on the logical-time semantics of Lingua Franca, it has a rather nice property.
The sensor fusion component can be designed to expect logically simultaneous inputs on even multiples of 50 ms and to expect only lidar inputs on odd multiples.
A simple change to the design in Fig.~\ref{fig:aeb} can accomplish much more sophisticated fault detection and handling with better timing properties.
We add a timer to the AutomaticEmergencyBraking system and change the reaction to the inputs to be triggered by the timer as suggested by the following pseudo code:
\begin{lstlisting}[numbers=none]
  timer t(0, 50 ms)
  reaction(t) lid, rad -> brake {=
    if (even multiple) {
      if (both inputs present) {
        // Normal fusion reaction.
      } else {
        // Fault handler
      }
    } else {
      if (only lidar is present) {
        // Normal lidar-only reaction
      } else {
        // Fault handler
      }
    }
  =} tardy {=
    // Fault handler
  =}
\end{lstlisting}
The second line uses an LF syntax to specify a reaction that is \textit{triggered} by the timer \texttt{t} and \textit{uses} inputs \texttt{lid} and \texttt{rad}, but is not triggered by them.
Invocation of this reaction will be delayed at most by the amount of the \textit{maxwait} regardless of whether inputs arrive on time.
Hence, missing inputs can be detected in bounded time, thereby enabling real-time reaction to faults.





A similar technique could be used in the AircraftDoor example of Fig.~\ref{fig:aircraft-door} (and many others).
The Cockpit could be equipped with a timer that periodically sends null messages (heartbeat messages).
A similar timer in the Door reactor could trigger a reaction that detects the \emph{absence} of such messages in time bounded at most by the \textit{maxwait}.
Such detection will occur even with total network failure as long as the local computer executing the Door is still operating.

\section{More Special Cases}
\label{sec:special}

A number of other commonly used patterns in distributed software can be realized using our \textit{maxwait} coordinator.

\subsection{Logical Execution Time (LET)}

Logical execution time (LET) is a principle for decoupling concurrent computations without introducing nondeterminism~\cite{Henzinger:01:Giotto,KirschSokolova:12:LET}.
The LET principle has recently been extended to distributed systems~\cite{Gemlau:21:LET} and used to improve cause-effect chains~\cite{Kohler:23:LET}.
The key idea is that an input to a computation that occurs at a certain logical time $t$ results in an output at logical time $t + E$, where $E$ is the logical execution time.
The semantics of Lingua Franca has been shown to be a generalization of synchronous language semantics and LET semantics~\cite{LeeLohstroh:22:LET}.

\begin{figure}[t]
  \centering
	  \includegraphics[width=\columnwidth]{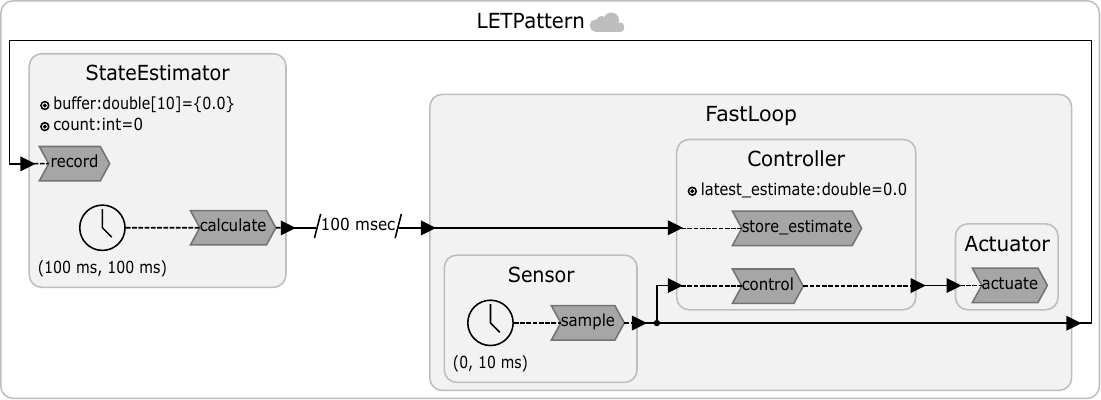}
  \caption{Pattern for logical execution time.}
  \label{fig:let}
\end{figure}

A typical pattern for LET is shown in Fig.~\ref{fig:let}.
Here, a FastLoop reactor operates a sensor, controller, and actuator at 100 Hz.
The sensor data is forwarded to a StateEstimator reactor, which records ten sensor values in a buffer and then calculates a state estimate.
Assume that this calculation is relatively time consuming.
The connection from the StateEstimator to the FastLoop has an \textit{after} delay, similar to the feedback path in Fig.~\ref{fig:optimistic-bank}.
This gives the calculate reaction a logical execution time of 100 ms.
That is, the output has a timestamp 100 ms larger than the triggering event.
This enables the calculation to proceed in the background in parallel with the faster operations in the FastLoop.

The StateEstimator has a \textit{maxwait} of \textit{forever}, which ensures that it always waits for 10 new input samples before starting the calculation.
The FastLoop reactor has \textit{maxwait} of zero, which works well if the calculation (plus communication overhead) takes less than the LET of 100 ms.
Every tenth firing of the Controller will find a new state estimate at its input.

It is typical in LET applications to assume that the physical time it takes to execute a computation is less than its logical execution time.
This implicit assumption can be checked at runtime using the \textit{maxwait} mechanism.
If we augment the Controller in Fig.~\ref{fig:let} with a timer with
(offset, period) = (200 ms, 100 ms), then we can add a reaction that checks for the presence of the expected state estimate.
If it is not present, then the timing assumption has been violated.
A fault handler needs to deal with this violation.

\subsection{Publish and Subscribe}

Pub-sub is popular even in situations where its intrinsic nondeterminism should be worrisome.
For example, ROS 2 (robot operating system) is widely used for robotics.
The underlying DDS (data distribution service) is used in many other cyber-physical applications.

In a pub-sub communication fabric, a node publishes data on a named topic, and other nodes subscribe to the topic.
Subscribers provide a callback function that is invoked when published messages have arrived.
If a subscriber subscribes to more than one topic, however, there is no semantics to the order in which messages are received, which leads to nondeterminism that can yield unexpected behaviors~\cite{BateniEtAl:23:Risk}.

Nevertheless, for some applications, such nondeterminism is not a problem (except that it complicates testing).
Fortunately, it is easy to use our \textit{maxwait} mechanism to achieve something similar to pub-sub.

In LF, the equivalent to a ``topic'' is an output port.
Any node can ``subscribe'' to that topic by simply connecting to that output port.
To achieve nondeterminism similar to pub-sub, simply set the \textit{maxwait} of the subscriber to zero (which happens to be the default).
Messages will be handled when they happen to arrive, just as with pub-sub.

The situation is slightly better than what is found in pub-sub frameworks such as ROS 2, however, because of the timestamps and clock synchronization provided by LF.
When a message arrives, even if it is tardy (an equal or later timestamp has already been processed), the intended tag of the message is available to the tardy handler.
Combined with clock synchronization, it becomes possible to infer much more information about the intended order than what is possible with ordinary pub-sub.

\subsection{Actors}

Frameworks such as Akka~\cite{RoestenburgEtAl:23:Akka} and CAF~\cite{
CharoussetEtAl:14:CAF} that support the Hewitt-Agha actor model of computation~\cite{Agha:86:Actors} are also quite popular despite their intrinsic nondeterminism.
In the actor model, a node sends messages to other nodes that it knows about, typically using an abstraction that looks like remote invocation of a message handler.
When a node's message handlers are invoked by remote nodes, they will be typically invoked in a mutually exclusive fashion, but there is no ordering semantics.
The order in which they are invoked will be determined by networking and scheduling timings that are not under the control of the framework.

Such behavior is again easy to emulate using LF with \textit{maxwait}.
In LF, a node (a reactor) does not send a message directly to another node, but rather sends a message via an output port to whatever other reactors are connected to that output port.
This is a small syntactic difference, but it has important consequences for modularity.
When designing a reactor, there is no need to know what other reactors it might later be connected to.
This modularity is not available with actors.

Nevertheless, as with pub-sub, simply setting \textit{maxwait} to zero, one can achieve semantics similar to actors with the added benefit that intended tags are available.

\subsection{Remote Procedure Calls}

\begin{figure}[t]
  \centering
	  \includegraphics[width=\columnwidth]{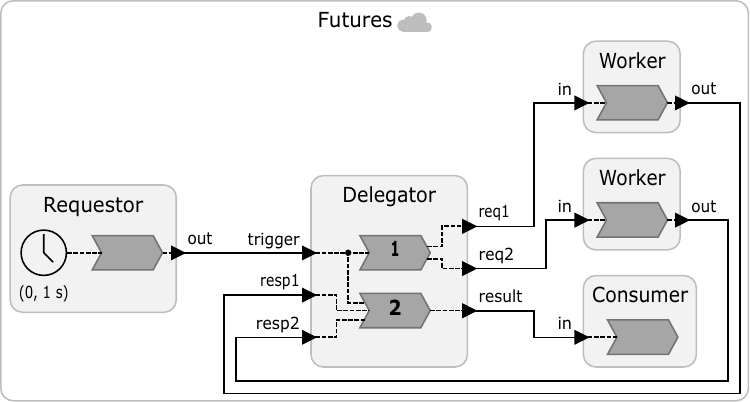}
  \caption{Pattern for futures in remote procedure calls.}
  \label{fig:futures}
\end{figure}

The \textit{maxwait} attribute prevents a reactor from advancing to a tag until either the inputs are known up to that tag or the timeout expires.
But there is a useful pattern where we want to advance to a tag even while some inputs remain unknown at that tag.
We then want to prevent any reactions that depend on those unknown inputs from executing until the inputs become known or a timeout expires.
Specifically, the pattern is where a reactor sends a request to a remote federate and then wants to wait for a response before proceeding. This is a common pattern in distributed systems known as remote procedure calling (RPC).

Consider the example shown in Fig~\ref{fig:futures}. The Requestor reactor issues requests for computation every one second. These requests are sent to the Delegator reactor, which offloads the computation to two Worker reactors and waits for their response. Once the workers have completed, the delegator aggregates and sends the result of the computation to a Consumer reactor.

The behavior we require in this example is that the Delegator reactor executes reaction 1 as soon as the trigger from the Requestor arrives, starting the two workers, but we want reaction 2 to be invoked only when the workers send back their response or after a timeout expires. All federates have \textit{maxwait} set to zero, which is the default value, and this allows the Delegator reactor to react immediately to the trigger from the requestor and execute reaction 1. However, \textit{maxwait} alone does not prevent immediate execution of reaction 2.

Our LF coordination mechanism features another attribute named \textit{absent\_after}. This is a property of a connection that specifies a timeout under which, after a reactor has advanced its current tag to $t$, if an input on that connection remains unknown when the timeout expires, then it will be assumed absent at tag $t$. With \textit{absent\_after}, when some inputs are unresolved at a tag $t$, we can delay the invocation of only those reactions that depend on those inputs, while still allowing the reactor advance to tag $t$ and execute reactions that do not depend on them.

In our example, the feedback connections from the workers to the Delegator reactor have \textit{absent\_after} set to 100 ms, which is done in LF with the following syntax:
\begin{lstlisting}[numbers=none]
  @absent_after(100 ms)
  w1.out -> d.resp1
  @absent_after(100 ms)
  w2.out -> d.resp2
\end{lstlisting}
This, combined with \textit{maxwait} set to zero, allows the Delegator reactor to immediately advance the tag when it receives a trigger from the Requestor reactor and invoke reaction 1, which starts the workers. Reaction 2 will be delayed until either both responses from the workers arrive, or the 100 ms timeout expires. A finite value for \textit{absent\_after} enables fault detection of the workers, as shown by the reaction 2 code:
\begin{lstlisting}[numbers=none,float,floatplacement=H!]
  reaction(trigger,resp1, resp2) -> result {=
    if (resp1->is_present && resp2->is_present) {
      lf_set(result, resp1->value+resp2->value);
    } else {
      // One or both of the responses is absent.
      // This is a failure, the result is zero.
      lf_set(result, 0);
      lf_print("Workers didn't respond on time");
    }
  =} tardy {=
    lf_print("Tardy response from workers");
  =}
\end{lstlisting}

If any of the responses is missing at the expiration of the timeout, a failure has occurred, and we signal it by setting the result to zero. Notice that including the trigger input from the Requestor reactor in the triggers of reaction 2 ensures that the reaction is invoked once for every trigger, even if both workers fail.
The \textit{absent\_after} attribute is similar to the safe-to-assume-absent (STAA) parameter, given by Donovan et al.~\cite{DonovanEtAl:25:ZDC}, which uses a centralized coordination strategy rather than our decentralized one.

In our example, if we instead set \textit{absent\_after} to \textit{forever}, reaction 2 will always wait for both inputs before executing. This realizes something like the ``futures'' concept of the RPC pattern, according to which a remote invocation immediately returns a future object, even if the remote response has not come yet. This permits performing other work within the caller while the callee works on the request. When the caller first uses the future object, it blocks until the remote server provides the response. Fig~\ref{fig:futures} implements exactly this pattern; any number of reactions could be put between reactions 1 and 2, thereby enabling computation to occur while waiting for the response from the Workers, just as with futures.



\section{Conclusion}

We have extended the Lingua Franca coordination language with a decentralized coordination mechanism that we call \textit{maxwait} that is surprisingly versatile.
It can realize many mechanisms that are used in distributed systems, and, most importantly, provides ways to control timing for time-sensitive systems.
It realizes conservative techniques that provide determinism at the expense of timing, optimistic techniques that provide timely responses at the expense of having to sometimes roll back, and balanced techniques that provide timing guarantees with measured risk of inconsistency. It also realizes popular patters such as logical execution time (LET), publish-and-subscribe, actors, and remote procedure calls with futures, all with added benefits that enable timely detection of anomalies and faults.


\bibliographystyle{plainurl}
\bibliography{references}

\end{document}